\documentclass[11pt,a4paper]{article}
\usepackage{jheppub} 
\usepackage{amsfonts,amssymb,latexsym,amsmath,mathrsfs,amsbsy}
\usepackage{bm}
\usepackage{slashed,bbold}
\usepackage[hhmmss]{datetime}
\usepackage{dsfont}

\newcommand{\mcB}{\mathcal B}

\newcommand{\mcV}{{\mathcal V}}

\def\fslash#1{#1\!\!\!/\,}

\def\fslashnabla{\nabla\!\!\!\!/\,}
\def\fslashnablabar{\overline{\nabla}\!\!\!\!/\,}

\usepackage[latin1]{inputenc}

\title{Chiral Torsional Effects in Anomalous Fluids in Thermal Equilibrium} 

\author[a]{Juan L. Ma\~nes,}
\author[a]{Manuel Valle,}
\author[b]{Miguel \'A. V\'azquez-Mozo}

\affiliation[a]{Departamento de F\'\i sica, 
Universidad del Pa\'is Vasco UPV/EHU, \\
Apartado 644,  48080 Bilbao, Spain}
\affiliation[b]{Departamento de F\'\i sica Fundamental, Universidad de Salamanca, \\
Plaza de la Merced s/n, 37008 Salamanca, Spain}

\emailAdd{wmpmapaj@lg.ehu.es}
\emailAdd{manuel.valle@ehu.es}
\emailAdd{Miguel.Vazquez-Mozo@cern.ch}

\abstract{
Using the similarity between spacetime torsion and axial gauge couplings,
we study torsional contributions to the equilibrium partition function in a stationary background. 
In the case of a charged fluid minimally coupled to torsion, we spot the existence of linear torsional
magnetic and vortical effects, while the axial-vector
current and the spin energy potential do not receive corrections in the torsion at linear order. 
The covariant energy-momentum tensor, on the other hand, does contain terms linear in the torsion tensor. 
The case of a two-flavor hadronic superfluid is also analyzed, and the torsional contributions
to the constitutive relations computed. 
Our results show the existence of a torsional electric chiral effect mediated by the charged pions. 
}

\arxivnumber{}

\begin{document}

\maketitle


\flushbottom


\section{Introduction}
\label{sec:intro}

The coupling of hydrodynamic systems to external sources through anomalous currents gives rise to a variety of
chiral transport effects~\cite{Son:2009tf,Fukushima:2008xe,Sadofyev:2010pr,Neiman:2011mj,Kirilin:2012mw,Fukushima:2012vr,Zakharov:2012vv,Landsteiner:2016led}. These have been shown to be 
relevant for the understanding of diverse physical phenomena, ranging
from condensed
matter to cosmology. Due to the topological nature of anomalies, the effective action 
for the long wavelength modes
 can be computed using 
differential geometry techniques,
from which the parity-violating terms in the fluid constitutive relations can be derived~\cite{Banerjee:2012iz,Jensen:2012jh,Jensen:2013kka,Jensen:2012kj,Jensen:2013rga,Haehl:2013hoa,Monteiro:2014wsa,Jain:2015jla,Banerjee:2015hra,Glorioso:2017lcn,Manes:2018llx,Manes:2019fyw}. 
By placing the system on a curved background stationary metric,
and performing dimensional reduction onto the compactified Euclidean time,
it is possible to incorporate physical effects such as vorticity and acceleration, 
which are sourced respectively by the background Kaluza-Klein (KK) gauge field and the gradient of the 
time component of the metric. 
Phenomena linked to the existence of mixed gauge-gravitational anomalies~\cite{Landsteiner:2011cp} 
have been subject to direct detection in the laboratory~\cite{Gooth:2017mbd}. 

Despite its absence in standard general relativity, torsion has been the focus of attention in physics for almost a century, 
since this geometrical notion was first introduced in the classical works
of \'Elie Cartan~\cite{Cartan:1922,Cartan:1923zea,Cartan:1924yea}. An obvious motivation for these investigations has been 
the possibility of our spacetime having a small albeit nonvanishing torsion, giving rise to 
new physics (see 
\cite{Hehl:1976kj,Shapiro:2001rz} for a review of different physical scenarios). 

Together with the prospects of spotting fundamental microscopic torsion in high-energy physics, 
torsional geometries provide a practical way of implementing physical effects 
in condensed matter physics. Focusing on the physics at large distances, 
the vectors defining the links at each node of the ion lattice
build up an effective dreibein whose geometry models lattice irregularities. For example, their curvature and torsion 
respectively implement lattice dislocations  
and disclinations~\cite{Hughes:2012vg}. In systems with linear dispersion relations, such
as the case of Weyl semimetals, this geometry provides a static nondynamical effective background  
on which fermions propagate~\cite{Parrikar:2014usa}.

Torsion is also known to have an interplay with chiral anomalies in quantum field 
theory~\cite{Zumino:1983ew,AlvarezGaume:1985ex,Bertlmann:1996xk,Fujikawa:2004cx}. 
The axial anomaly receives a contribution given by the so-called Nieh-Yan 
term \cite{Nieh:1981ww,Nieh:1981xk}, which comes from a bubble diagram with two axial-vector current
insertions and is quadratically divergent. As a consequence, the coefficient of the Nieh-Yan term
depends of the square of the cutoff, or any other relevant UV scale of the theory~\cite{Chandia:1997hu}. 
The role of this term in condensed matter physics has been explored in 
a number of works (see, for example,~\cite{Hidaka:2012rj,Hughes:2012vg,Parrikar:2014usa,Nissinen:2019kld,Nissinen:2019mkw}).

In the context of chiral fluids, the interest in torsion arose in 
connection with the study of Hall 
transport~\cite{Son:2007ny,Haehl:2013kra,Son:2013rqa,Geracie:2014mta}
(see~\cite{Hoyos:2014pba} for a comprehensive review). The relation between the Hall viscosity
and the mean orbital spin per particle suggested a connection with the spin current, which is sourced by 
the background torsion. In the relativistic setup, this link 
was further studied in Ref.~\cite{Valle:2015hfa} by 
a first principles calculation of the effective action and constitutive relations of a fermion gas on a $(2+1)$-dimensional
spacetime with torsion. 

The issue of torsional transport effects in four dimensions has also received some attention lately~\cite{Khaidukov:2018oat}. Besides the sector associated to the Nieh-Yan anomaly\footnote{It 
has recently been proposed, however, that there is no genuine torsional chiral 
dissipationless transport in that sector~\cite{Ferreiros:2020uda}.}, 
the partition function contains other contributions 
which are induced by the triangle diagrams associated with the effective axial-vector field encoding
the antisymmetric part of the torsion. 
In Ref.~\cite{Imaki:2020csc}, transport phenomena induced by  
torsion were studied, both at zero and finite temperature. The existence of a chiral magnetic and electric effects 
was found, resulting from fermions minimally coupled to 
the antisymmetric part of the torsion tensor, which can be recast in terms of its dual vector field. 
This external source couples to the fermionic singlet axial-vector current, which is affected by a 't Hooft anomaly. 
Torsional contributions to spin transport were also recently studied in~\cite{Gallegos:2020otk}.

The effects associated with the 't Hooft anomaly of the gauge field dual to the antisymmetric part of the
torsion
can be readily computed using the standard differential geometry 
methods employed in the analysis of anomalous fluids.
In the present work, we apply the techniques developed 
in~\cite{Manes:2018llx,Manes:2019fyw} to carry out a  
study of the linear effects of torsion in hydrodynamics, 
with and without Nambu-Goldstone bosons. 
For a charged fluid
coupled to an external electromagnetic field, 
we verify the existence  
of torsional magnetic and vortical effects. The axial-vector current, on the other hand, does not contain any corrections
linear in the torsion. This is also the case for the covariant spin energy potential, which is written in terms
of the covariant axial-vector current. The components of the covariant energy-momentum tensor
can be also expressed in terms of the covariant axial-vector currents, but in this case the coefficients depend 
linearly on the torsion tensor. 
Once written in terms of the torsion, they give rise to new torsion-induced 
contributions to the constitutive relations from where the corresponding transport coefficients can be obtained.

After analyzing the Abelian case, we focus our attention on the case of a two-flavor hadronic superfluid in the
presence of torsion, in the phase in which chiral symmetry $\mbox{U(2)}_{L}\times\mbox{U(2)}_{R}$ 
is spontaneously broken to its vector subgroups. 
We compute the corrections
to the covariant currents and transport coefficients linear in the torsion tensor and find the existence of 
a torsional chiral electric effect mediated by the two charged pions. The chiral separation effects found in~\cite{Manes:2019fyw},
on the other hand, do not receive any contributions linear in the torsion.

The present article is organized as follows.
Section~\ref{sec:hydro_torsion} is devoted to the analysis of the equilibrium partition function of a charged 
plasma in the presence of torsion, including the computation of the covariant currents. This models is further elaborated in 
Section~\ref{sec:spin+emtensor} with the calculation of linear torsional contributions to the spin energy potential and
the energy-momentum tensor. In Section~\ref{sec:hadronic}, after a brief discussion
of the linear coupling of Nambu-Goldstone bosons to torsion in the Abelian case, we compute the linear torsional corrections to 
the constitutive relations of a two-flavor hadronic superfluid. 
Finally, our findings are summarized in Section~\ref{sec:conclusions}. To make our presentation more self-contained, we 
review in Appendix~\ref{sec:torsion} some basic facts about geometric torsion, while in Appendix~\ref{app:identities} 
we list some expressions of Ref.~\cite{Manes:2019fyw} relevant to our discussion.

\section{Equilibrium partition function and covariant currents with background torsion}
\label{sec:hydro_torsion}

We begin with the discussion of the dynamics of massless Dirac fermions propagating on 
a spacetime with torsion\footnote{The basics of geometric torsion, as well as the notation used in the 
following, are summarized in Appendix~\ref{sec:torsion}.}. 
The action of a massless Dirac spinor minimally coupled to gravity
can be written as~\cite{Hehl:1976kj,Ortin:2015hya,Freedman:2012zz}
\begin{align}
S={1\over 2}\int d^{4}x\,(\det{e})\Big(\overline{\psi}\overrightarrow{\fslashnabla}\psi
-\overline{\psi}\overleftarrow{\fslashnabla}\psi\Big),
\label{eqw:action_mincoupgrav}
\end{align}
where the left and right covariant derivatives inside the integral are defined respectively by
\begin{align}
\overrightarrow{\fslashnabla}\psi&=\gamma^{A}e_{A}^{\,\,\,\mu}\partial_{\mu}\psi+{1\over 4}\gamma^{A}
\gamma^{[B}\gamma^{C]}\omega_{BCA}\psi,\nonumber \\[0.2cm]
\overline{\psi}\overleftarrow{\fslashnabla}&=e_{A}^{\,\,\,\mu}\partial_{\mu}\overline{\psi}\gamma^{A}
-{1\over 4}\overline{\psi}\gamma^{[B}\gamma^{C]}\gamma^{A}\omega_{BCA},
\label{eq:right_left_derivs}
\end{align}
and the Dirac matrices verify the Minkowskian Clifford algebra $\{\gamma^{A},\gamma^{B}\}=2\eta^{AB}\mathbb{1}$. 
Writing the full spin connection in terms of the auxiliary torsionless Levi-Civita connection and the contorsion tensor 
as~$\omega^{A}_{\,\,\,BC}=\overline{\omega}^{A}_{\,\,\,BC}+\kappa^{A}_{\,\,\,BC}$, we get the following
expression of the left covariant derivative in terms
of its Levi-Civita counterpart
\begin{align}
\overrightarrow{\fslashnabla}\psi&=\overrightarrow{\fslashnablabar}\psi
+{1\over 4}\gamma^{C}\gamma^{[A}\gamma^{B]}\kappa_{ABC}\psi \nonumber \\[0.2cm]
&=\overrightarrow{\fslashnablabar}\psi
+{1\over 4}\Big(\gamma^{B}\eta^{AC}-\gamma^{A}\eta^{BC}+i\epsilon^{ABCD}\gamma_{D}\gamma_{5}\Big)\kappa_{ABC}\psi, 
\end{align}
where we indicate by a bar all geometric quantities referred to
the Levi-Civita connection and have used the gamma matrices identity
\begin{align}
\gamma^{A}\gamma^{[B}\gamma^{C]}=\gamma^{C}\eta^{AB}-\gamma^{B}\eta^{AC}+i\epsilon^{ABCD}\gamma_{D}\gamma_{5}.
\end{align}
A similar calculation for the right covariant derivative in Eq.~\eqref{eq:right_left_derivs} gives
\begin{align}
\overline{\psi}\overleftarrow{\fslashnabla}&=\overline{\psi}\overleftarrow{\fslashnablabar}-{1\over 4}\overline{\psi}\gamma^{[A}\gamma^{B]}\gamma^{C}
\kappa_{ACB} \nonumber \\[0.2cm]
&=\overline{\psi}\overleftarrow{\fslashnablabar}-{1\over 4}\overline{\psi}\Big(\gamma^{A}\eta^{BC}-\gamma^{B}\eta^{AC}+i\epsilon^{ABCD}\gamma_{D}
\gamma_{5}\Big)\kappa_{ABC}.
\end{align}
Plugging these results into the action~\eqref{eqw:action_mincoupgrav}, we arrive at the expression
\begin{align}
S&={1\over 2}\int d^{4}x\,(\det{e})\left[\overline{\psi}\overrightarrow{\fslashnablabar}\psi
+{1\over 4}\overline{\psi}\Big(\gamma^{B}\eta^{AC}-\gamma^{A}\eta^{BC}+i\epsilon^{ABCD}\gamma_{D}\gamma_{5}\Big)\psi\kappa_{ABC}
\right. \nonumber \\[0.2cm]
&\left.-\overline{\psi}\overleftarrow{\fslashnablabar}\psi+{1\over 4}\overline{\psi}\Big(\gamma^{A}\eta^{BC}-\gamma^{B}\eta^{AC}+i\epsilon^{ABCD}\gamma_{D}
\gamma_{5}\Big)\psi\kappa_{ABC}\right].
\label{eq:fermion_coupling_all_terms}
\end{align}
The important point here is that 
the term proportional to $(\gamma^{B}\eta^{AC}-\gamma^{A}\eta^{BC})\kappa_{ABC}$, which contains the
symmetric components of the contorsion in the two last indices, cancels out. This means that fermions only 
couple to 
its antisymmetric piece, $\kappa^{A}_{\,\,\,\,[BC]}$, which as shown in Appendix~\ref{sec:torsion} [see Eq.~\eqref{eq:antisymmetric_part_kappa}] is given by the
components of the torsion tensor
\begin{align}
S&={1\over 2}\int d^{4}x\,(\det{e})\left(\overline{\psi}\overrightarrow{\fslashnablabar}\psi-\overline{\psi}\overleftarrow{\fslashnablabar}\psi
+{i\over 4}\overline{\psi}\gamma^{D}\gamma_{5}\psi\epsilon_{DA}^{\,\,\,\,\,\,\,\,\,BC}T^{A}_{\,\,\,\,\,BC} 
\right).
\end{align} 
This form of the action suggests the introduction of the effective vector field
\begin{align}
\mathcal{S}_{A}=-{1\over 8}\epsilon_{AB}^{\,\,\,\,\,\,\,\,\,\,CD}T^{B}_{\,\,\,\,\,\,CD},
\label{eq:SepsilonT_comp}
\end{align}
to write
\begin{align}
S&={1\over 2}\int d^{4}x\,(\det{e})\left(\overline{\psi}\overrightarrow{\fslashnablabar}\psi-\overline{\psi}\overleftarrow{\fslashnablabar}\psi
-2i\overline{\psi}\gamma^{A}\gamma_{5}\psi\mathcal{S}_{A}\right).
\label{eq:action_effective_axial_screw}
\end{align}
Thus, the whole effect of background torsion on the dynamics of the fermion
is codified through its axial-vector coupling to an external effective gauge field, 
which, following Ref.~\cite{Imaki:2020csc}, we call screw torsion. 
The minimal coupling to gravity selects only one among all possible dimension-four operators coupling
Dirac fermions to torsion~\cite{Kostelecky:2007kx}. 

Before proceeding any further, some clarification on the action~\eqref{eqw:action_mincoupgrav} is in order.
At face value, the theory it describes seems to be equivalent to that of a Dirac fermion axially coupled to
an external gauge field. The crucial difference, however, is that this external gauge 
field~$\mathcal{S}$ 
is a ``composite'' expressed as the Hodge dual of the antisymmetric part of the torsion components,
which is the ``fundamental'' external source. This is important, because once expressed in a coordinate basis
the components of this gauge field depend not only on the torsion, but on the metric tensor as well. 
As a result, the energy-momentum tensor and the spin energy potential of the theory include contributions 
that would be absent in the 
theory of a ``fundamental'' gauge field axially coupled to a Dirac fermion (see Sec.~~\ref{sec:spin+emtensor}). 
These new terms are associated with novel transport coefficients in the constitutive relations for the 
corresponding energy-momentum and spin covariant currents.

In addition to the couplings shown in Eq.~\eqref{eqw:action_mincoupgrav}, the authors of
Ref.~\cite{Imaki:2020csc} considered an additional coupling of the Dirac fermion to an external
Abelian vector gauge field. This so-called edge torsion vector field is proportional to the torsion vector 
$T^{B}_{\,\,\,\,BA}$ and mixes for many practical purposes with the electromagnetic field. 
In what follows we stick to the minimal coupling prescription~\eqref{eqw:action_mincoupgrav} and
only consider the coupling to the screw torsion, in the understanding that in all our results the edge torsion
would be reabsorbed by a shift in the physical electromagnetic field.

\paragraph{Remarks on gauge invariance.}
We have seen how the action of a Dirac fermion minimally coupled to gravity only depends on the 
fully antisymmetric components of the torsion tensor $\mathcal{T}_{ABC}\equiv T_{[ABC]}$, 
which can be used to define the three-form\footnote{Unlike in Refs.~\cite{Manes:2018llx,Manes:2019fyw}, here
no $-i$ is factored out of the components of differential forms.}
\begin{align}
\mathcal{T}={1\over 3!}\mathcal{T}_{ABC}e^{A}e^{B}e^{C}.
\label{eq:antisymm_torsion_form}
\end{align}
Its four independent components are encoded in the screw-torsion according to Eq.~\eqref{eq:SepsilonT_comp}, which 
can be written using the Hodge star operator as
\begin{align}
\mathcal{S}_{\mu}=-{1\over 8}\epsilon_{\mu\nu}^{\,\,\,\,\,\,\,\alpha\beta}\mathcal{T}^{\nu}_{\,\,\,\,\,\alpha\beta}
 \hspace*{1cm} \Longrightarrow \hspace*{1cm}
\mathcal{S}={3\over 4}\star \mathcal{T}.
\label{eq:SvsT_forms}
\end{align}
The field strength of the Abelian screw torsion, $\mathcal{F}_{\mathcal{S}}=d\mathcal{S}$, can be written then as
\begin{align}
\mathcal{F}_{\mathcal{S}}&={3\over 4}d\star\mathcal{T}=-{3\over 4}\star\delta\mathcal{T},
\label{eq:screw_field_strength_dform}
\end{align}
where $\delta\equiv-\star d\star$ denotes the codifferential acting on a three-form. In components, this equation reads
\begin{align}
\mathcal{S}_{\mu\nu}=-{3\over 8}\overline{\nabla}_{\sigma}\mathcal{T}^{\sigma}_{\,\,\,\,\,\alpha\beta}
\epsilon^{\alpha\beta}_{\,\,\,\,\,\,\,\mu\nu}.
\end{align}

Looking at Eq.~\eqref{eq:SvsT_forms} above, we see that the gauge variation of the screw torsion vector field $\mathcal{S}$
by an exact one-form, $\mathcal{S}\rightarrow\mathcal{S}+d\alpha$, corresponds to the following transformation 
of the torsion three-form~$\mathcal{T}$
\begin{align}
\mathcal{T}\longrightarrow \mathcal{T}+\delta\beta, 
\label{eq:Tform_gauge}
\end{align}
with $\beta\sim \star\alpha$ a four-form. The nihilpotency of the codifferential, $\delta^{2}=0$, guarantees the 
gauge invariance of the screw torsion field strength~\eqref{eq:screw_field_strength_dform}. 

\paragraph{The equilibrium partition function.}
In the context of hydrodynamics, 
the coupling of the background torsion to the microscopic fermionic degrees of freedom gives the prescription for the 
construction of the effective functional describing the long-range excitations of a fluid \cite{Banerjee:2012iz,Jensen:2012jh}. 
Here we are going to employ the differential geometry methods introduced in \cite{Manes:2018llx,Manes:2019fyw} to build the
equilibrium partition function for fluids with torsion. In the following, we study the case
of a fluid coupled to 
an external Abelian vector source in the presence of background torsion on a generic static background
geometry. 
After this, the more general case of a two flavor hadronic 
(super)fluid will be analyzed in Section \ref{sec:hadronic}.

Besides its coupling to torsion through $\mathcal{S}$, we also assume that 
the microscopic fermionic degrees of freedom are coupled to an external vector Abelian gauge field 
$\mathcal{V}$, which remains anomaly-free and will be eventually associated to the electromagnetic field. 
In four dimensions, the (nonlocal) anomalous part of the effective action can be computed in terms of the 
Chern-Simons form (see, for example,~\cite{Manes:2018llx}). Keeping only terms linear in the torsion, we have
\begin{align}
\widetilde{\omega}^{0}_{5}(\mathcal{S},\mathcal{F}_{V},\mathcal{F}_{\mathcal{S}})&=6\mathcal{S}\mathcal{F}_{V}^{2},
\end{align}
where $\mathcal{F}_{V}=d\mathcal{V}$ is the vector field strength. 
This gives the Bardeen form of the anomaly, which explicitly preserves vector gauge transformations. 
The properly normalized Chern-Simons nonlocal
effective action encoding the linear effects of torsion is then given by
\begin{align}
\Gamma[\mathcal{V},\mathcal{S}]_{\rm CS}={1\over 4\pi^{2}}\int\limits_{D_{5}}
\mathcal{S}\mathcal{F}_{V}^{2},
\end{align}
where $D_{5}$ is a five-dimensional manifold whose boundary is identified with the Euclidean four-dimensional physical spacetime. 
To compute the equilibrium partition function from the Chern-Simons effective action, 
we take the metric of the four-dimensional spacetime $\partial D_{5}$ to be the generic static line element
\begin{align}
ds^{2}=-e^{2\sigma(\mathbf{x})}\Big[dx^{0}+a_{i}(\mathbf{x})dx^{i}\Big]^{2}+g_{ij}(\mathbf{x})dx^{i}dx^{j},
\label{eq:static_line_element}
\end{align}
and take all fields to be independent of $x^{0}$.
We implement dimensional reduction onto the compatified Euclidean time by setting 
$D_{5}=S^{1}\times D_{4}$, where the length of the $S^{1}$ 
equals the inverse of the equilibrium temperature $T_{0}$. Vector fields are then written in terms of components that remain invariant
under KK transformations~\cite{Manes:2019fyw}, acting according to
$x^{0}\rightarrow x^{0}+\phi$ and $a_{i}\rightarrow a_{i}-\partial_{i}\phi$,
\begin{align}
\mathcal{V}&\equiv \mathcal{V}_{\mu}dx^{\mu}=\boldsymbol{V}-e^{-\sigma}\mathcal{V}_{0}u, \nonumber \\[0.2cm]
\mathcal{S}&\equiv \mathcal{S}_{\mu}dx^{\mu}=\boldsymbol{S}-e^{-\sigma}\mathcal{S}_{0}u,
\label{eq:VS_def}
\end{align}
where we have introduced the four-velocity one-form $u$ given by
\begin{align}
u=-e^{\sigma}\big(dx^{0}+a_{i}dx^{i}\big)\equiv -e^{\sigma}\big(dx^{0}+a\big),
\label{eq:velocity_one-form}
\end{align}
and the KK-invariant spatial one-forms are defined by
\begin{align}
\boldsymbol{V}&=\Big(\mathcal{V}_{i}-\mathcal{V}_{0}a_{i}\Big)dx^{i}\equiv V_{i}dx^{i}, \nonumber \\[0.2cm]
\boldsymbol{S}&=\Big(\mathcal{S}_{i}-\mathcal{S}_{0}a_{i}\Big)dx^{i}\equiv S_{i}dx^{i}.
\end{align}
A similar electric-magnetic decomposition can be written for the vector field strength
\begin{align}
\mathcal{F}_{V}&\equiv \boldsymbol{B}+u\boldsymbol{E}
\nonumber \\[0.2cm]
&=d\boldsymbol{V}-d\big(e^{-\sigma}u\big)\mathcal{V}_{0}+ue^{-\sigma}d\mathcal{V}_{0} 
\label{eq:emdecomFV} \\[0.2cm]
&=\Big(\boldsymbol{F}_{V}+\mathcal{V}_{0}da\Big)+ue^{-\sigma}d\mathcal{V}_{0}, 
\nonumber
\end{align}
where $\boldsymbol{F}_{V}\equiv d\boldsymbol{V}$
and we have used that $d(e^{-\sigma}u)=-da$. The electric $\boldsymbol{E}$ 
and magnetic $\boldsymbol{B}$ components of the field strength will be later identified with the electric and magnetic 
fields [cf. \eqref{eq:electric_magnetic_fields_def}].
The equivalent expression for the field strength associated
to the screw torsion reads
\begin{align}
\mathcal{F}_{S}&\equiv \boldsymbol{B}_{S}+u\boldsymbol{E}_{S} \nonumber \\[0.2cm]
&=\Big(\boldsymbol{F}_{S}+\mathcal{S}_{0}da\Big)+ue^{-\sigma}d\mathcal{S}_{0},
\label{eq:emdecomFS} 
\end{align}
with $\boldsymbol{F}_{S}=d\boldsymbol{S}$.

Finally, we implement the dimensional reduction on the 
three-form~\eqref{eq:antisymm_torsion_form} as well by decomposing it as
\begin{align}
\mathcal{T}=\mathcal{T}_{B}+u\mathcal{T}_{E}.
\label{eq:T_electric+magnetic}
\end{align}
In a coordinate basis, the 
electric and magnetic components are respectively given by
\begin{align}
\mathcal{T}_{E}&={1\over 3!}e^{-\sigma}\Big[2g_{i\ell}T^{\ell}_{\,\,\,0j}+e^{2\sigma}\Big(T^{0}_{\,\,\,ij}
-2a_{i}T^{0}_{\,\,\,0j}\Big)+e^{2\sigma}a_{\ell}\Big(T^{\ell}_{\,\,\,ij}-2a_{i}T^{\ell}_{\,\,\,0j}\Big)\Big]
dx^{j}dx^{k}, \nonumber \\[0.2cm]
\mathcal{T}_{B}&={1\over 3!}g_{j\ell}\Big(T^{\ell}_{\,\,\,\,kn}-2a_{k}T^{\ell}_{\,\,\,\,0n}\Big)
dx^{j}dx^{k}dx^{n}.
\label{eq:T_electric_magnetic_decomp}
\end{align}
Being a four-form, the gauge function $\beta$ in Eq.~\eqref{eq:Tform_gauge} does not have any magnetic component,
$\beta=u\beta_{E}$. As a consequence, only the electric part of $\mathcal{T}$ transforms under~\eqref{eq:Tform_gauge}
\begin{align}
\mathcal{T}_{E}&\longrightarrow \mathcal{T}_{E}+\delta_{\perp}\beta_{E}, \nonumber \\[0.2cm]
\mathcal{T}_{B}&\longrightarrow \mathcal{T}_{B},
\label{eq:T_invariance_electric_magnetic}
\end{align}
where $\delta_{\perp}=*d*$, with $*$ the three-dimensional Hodge dual (not to be confused with its 
four-dimensional couterpart denoted by $\star$). Since under four-dimensional Hodge duality the electric
and magnetic components interchange 
\begin{align}
\star\mathcal{T}=-*\mathcal{T}_{E}-u*\mathcal{T}_{B},
\end{align}
we find from Eqs.~\eqref{eq:SvsT_forms} and~\eqref{eq:VS_def}
\begin{align}
\mathcal{S}_{0}&={3\over 4}e^{\sigma}*\mathcal{T}_{B}, \nonumber \\[0.2cm]
\boldsymbol{S}&=-{3\over 4}*\mathcal{T}_{E}.
\end{align}
We see that the gauge invariance of $\mathcal{T}_{B}$ implies the same property for $\mathcal{S}_{0}$, whereas
$\boldsymbol{S}$ undergoes the standard gauge transformation generated by the zero-form $*\beta_{E}$. 
Using in addition Eq.~\eqref{eq:T_electric_magnetic_decomp}, we can write
the screw torsion field in terms of the 
components of the torsion tensor as
\begin{align}
\mathcal{S}_{0}&={1\over 8}e^{\sigma}\epsilon^{ijk}g_{i\ell}\Big(T^{\ell}_{\,\,\,\,jk}-2a_{j}T^{\ell}_{\,\,\,\,0k}\Big), 
\label{eq:S'svsT's}\\[0.2cm]
\boldsymbol{S}&=-{1\over 8}e^{-\sigma}\epsilon^{ijk}
g_{mi}\Big[2g_{\ell j}T^{\ell}_{\,\,\,0k}+e^{2\sigma}a_{\ell}\Big(T^{\ell}_{\,\,\,jk}
-2a_{j}T^{\ell}_{\,\,\,0k}\Big)
+e^{2\sigma}\Big(T^{0}_{\,\,\,jk}-2a_{j}T^{0}_{\,\,\,0k}\Big)
\Big]dx^{m}.
\nonumber
\end{align}
This dependence of the screw torsion gauge field on the metric and 
torsion components is what distinguishes our theory from that of a Dirac fermion coupled to 
an external gauge field through the axial-vector current. As already pointed out, this
has important consequences for the constitutive relations of the energy-momentum and spin currents.

Having arrived at this parametrization of the effective screw torsion, we
proceed to compute the terms in the 
effective action induced by the 't Hooft anomaly affecting the gauge
invariance~\eqref{eq:T_invariance_electric_magnetic} and coming from triangle diagrams\footnote{As stated above,
in this work we do not consider the sector associated with the Nieh-Yan anomaly.}.
As shown in \cite{Jensen:2013kka,Manes:2018llx}, the dimensionally-reduced effective action splits into a local anomalous and a nonlocal invariant
piece, respectively given by~\cite{Manes:2018llx}
\begin{align}
W[\mathcal{V}_{0},\mathcal{S}_{0},\boldsymbol{V},\boldsymbol{S},da]_{\rm anom}&={1\over 4\pi^{2}T_{0}}\int\limits_{S^{3}}
\Big(2\mathcal{V}_{0}\boldsymbol{F}_{V}
+da\mathcal{V}_{0}^{2}\Big)\boldsymbol{S},
\label{eq:anomalous_effaction}
\end{align}
and
\begin{align}
W[\mathcal{V}_{0},\mathcal{S}_{0},\boldsymbol{V},\boldsymbol{S},da]_{\rm inv}&={1\over 4\pi^{2}T_{0}}\int\limits_{D_{4}}\Big[
\mathcal{S}_{0}\boldsymbol{F}_{V}^{2}+2\mathcal{V}_{0}\boldsymbol{F}_{V}\boldsymbol{F}_{S}
\\[0.2cm]
&+da\mathcal{V}_{0}\Big(\mathcal{V}_{0}\boldsymbol{F}_{S}+2\mathcal{S}_{0}\boldsymbol{F}_{V}\Big)
+(da)^{2}\mathcal{S}_{0}\mathcal{V}_{0}^{2}\Big].
\label{eq:invariant_effaction}
\nonumber
\end{align}
In the second expression, we have introduced the components of the field strength associated with $\boldsymbol{S}$
\begin{align}
\boldsymbol{F}_{S}&= d\boldsymbol{S}\equiv {1\over 2}S_{ij}dx^{i}dx^{j}.
\end{align}

The covariant currents can be now computed from the invariant part of the partition function~\cite{Jensen:2013kka,Manes:2018llx,Manes:2019fyw}. 
We begin with the one associated with the vector current
\begin{align}
\langle\boldsymbol{J}_{V}\rangle_{\rm cov}&=T_{0}{\delta\over\delta \boldsymbol{F}_{V}}W[\mathcal{V}_{0},\mathcal{S}_{0},\boldsymbol{V},\boldsymbol{S}]_{\rm inv}
\nonumber \\[0.2cm]
&={1\over 2\pi^{2}}\Big[\mathcal{S}_{0}\Big(\boldsymbol{F}_{V}+da\mathcal{V}_{0}\Big)
+\mathcal{V}_{0}\boldsymbol{F}_{S}\Big],
\label{eq:abelian_JVcov}
\end{align}
whereas for the one corresponding to the torsional axial-vector gauge field, the result is
\begin{align}
\langle\boldsymbol{J}_{S}\rangle_{\rm cov}&=T_{0}{\delta\over\delta \boldsymbol{F}_{S}}W[\mathcal{V}_{0},\mathcal{S}_{0},\boldsymbol{V},\boldsymbol{S}]_{\rm inv}
\nonumber \\[0.2cm]
&={1\over 2\pi^{2}}\left(\mathcal{V}_{0}\boldsymbol{F}_{V}
+{1\over 2}da\mathcal{V}_{0}^{2}\right).
\label{eq:abelian_JScov}
\end{align}
In both cases, the zero components vanish, $\langle \mathcal{J}_{V0}\rangle_{\rm cov}=\langle
\mathcal{J}_{S0}\rangle_{\rm cov}=0$. We observe that, 
unlike the vector current, $\langle\boldsymbol{J}_{S}\rangle_{\rm cov}$
does not pick any linear dependence on the torsion. These covariant currents will be very relevant in the following.

In addition, the corresponding Bardeen-Zumino (BZ) currents are given by~\cite{Manes:2018llx}
\begin{align}
\langle\mathcal{J}_{V}\rangle_{\rm BZ}&={1\over 2\pi^{2}}\mathcal{S}\mathcal{F}_{V}, \nonumber \\[0.2cm]
\langle\mathcal{J}_{S}\rangle_{\rm BZ}&={1\over 6\pi^{2}}\mathcal{S}\mathcal{F}_{S}.
\end{align}
The second equation is quadratic in the screw torsion field, so that the BZ current does
not have any correction at linear order in the torsion. For the BZ vector current, on the other hand, 
after dimensional reduction we find
\begin{align}
\langle\mathcal{J}_{V0}\rangle_{\rm BZ}
&={1\over 2\pi^{2}}\Big(\boldsymbol{F}_{V}+\mathcal{V}_{0}da\Big)\boldsymbol{S}, \nonumber \\[0.2cm]
\langle\boldsymbol{J}_{V}\rangle_{\rm BZ}&=
{1\over 2\pi^{2}}
\Big[\mathcal{S}_{0}\Big(\boldsymbol{F}_{V}+\mathcal{V}_{0}da\Big)+\boldsymbol{S}d\mathcal{V}_{0}\Big].
\label{eq:BZ_currents}
\end{align}
Notice that only the magnetic part of the vector field strength enters in these expressions~[see Eq.~\eqref{eq:emdecomFV}].

To connect with physics, we need to identify the electric and magnetic fields. These are naturally given 
by the electric and magnetic parts in the decomposition~\eqref{eq:emdecomFV} of the vector field strength. 
Thus, we have\footnote{Our definition of the physical magnetic field follows~\cite{Jensen:2013kka} and
differs from the one
used in~\cite{Manes:2018llx,Manes:2019fyw}. Expressions in these two references can be adapted to our definition
by the replacement $\mathbb{B}^{i}\rightarrow \mathbb{B}^{i}+2\mu\omega^{i}$.}
\begin{align}
e\mathbb{E}_{i}&=e^{-\sigma}\partial_{i}\mathcal{V}_{0}, \nonumber \\[0.2cm]
e\mathbb{B}^{i}&=\epsilon^{ijk}\Big(\partial_{j}V_{k}+\mathcal{V}_{0}\partial_{j}a_{k}\Big),
\label{eq:electric_magnetic_fields_def}
\end{align}
with $e$ the elementary electric charge.
We also define the screw magnetic field as the three-dimensional Hodge dual of the
 magnetic component
of $\boldsymbol{B}_{S}$ in~\eqref{eq:emdecomFS}
\begin{align}
B^{i}_{S}=\epsilon^{ijk}\Big(\partial_{j}S_{k}+\mathcal{S}_{0}\partial_{j}a_{k}\Big).
\label{eq:BsDef}
\end{align}
In addition, we introduce the vorticity vector
\begin{align}
\omega^{i}&=-{1\over 2}e^{\sigma}\epsilon^{ijk}\partial_{j}a_{k},
\label{eq:magnetic+vorticity_fields}
\end{align}
and the electromagnetic and torsional chemical potentials
\begin{align}
\mu&\equiv e^{-\sigma}\mathcal{V}_{0}, \nonumber \\[0.2cm]
\mu_{S}&\equiv e^{-\sigma}\mathcal{S}_{0}.
\label{eq:chemical_potentials}
\end{align}
With all these definitions in hand, we take the Hodge duals of the covariant currents \eqref{eq:abelian_JVcov} and~\eqref{eq:abelian_JScov} to write the electromagnetic and axial-vector covariant current components
\begin{align}
\langle J^{i}_{\rm em}\rangle_{\rm cov}&
={1\over 2\pi^{2}}\Big(\mu_{S}e\mathbb{B}^{i}+\mu B_{S}^{i}+2\mu\mu_{S}\omega^{i}\Big), 
\nonumber \\[0.2cm]
\langle J^{i}_{5}\rangle_{\rm cov}&={\mu\over 2\pi^{2}}\Big(e\mathbb{B}^{i}
+\mu\omega^{i}\Big).
\label{eq:torsional_effects_vmodel}
\end{align}
As argued in~\cite{Jensen:2013kka,Haehl:2013hoa,Manes:2018llx,Manes:2019fyw}, covariant currents are the ones relevant for the analysis of transport. Thus,  
the conclusion to be extracted from our result is the existence of torsional 
chiral magnetic and vortical effects, both mediated by the torsional chemical potentials, together with a nondissipative
transport of charge driven by the screw torsion magnetic field~\eqref{eq:BsDef}. 
The right-hand side of the second expression, on the other hand, gives 
the standard result for the axial-vector current of an Abelian anomalous fluid, corresponding to the
magnetic and vortical separation effect without any torsional corrections at linear order. Notice that here
we have identified the current $J^{i}_{S}$
with the axial-vector current, $J^{i}_{5}$.

We also give the components of the vector BZ currents by taking the dual of the expressions in Eq.~\eqref{eq:BZ_currents} and recasting them in term of electric and magnetic fields
\begin{align}
\langle J_{{\rm em}\,0}\rangle_{\rm BZ}
&={e\over 2\pi^{2}}e^{\sigma}\mathbb{B}^{i}S_{i}, \nonumber \\[0.2cm]
\langle J_{\rm em}^{i}\rangle_{\rm BZ}&=
{e\over 2\pi^{2}}\Big(\mu_{S}\mathbb{B}^{i}+\epsilon^{ijk}S_{j}\mathbb{E}_{k}\Big).
\label{eq:BZemcurrent}
\end{align}
This BZ current was identified in Ref.~\cite{Imaki:2020csc} as the electromagnetic current associated
with charge transport in flat spacetime. It differs however
from the covariant current computed above by the consistent current
\begin{align}
\langle J_{\rm em}^{i}\rangle_{\rm cons}\equiv
\langle J_{{\rm em}}^{i}\rangle_{\rm cov}-\langle J_{{\rm em}}^{i}\rangle_{\rm BZ}
={1\over 2\pi^{2}}\partial_{j}\big(\mathcal{V}_{0}\epsilon^{ijk}S_{k}\big),
\end{align}
whose divergence vanish identically due to the presence of the Levi-Civita tensor.

\section{The spin energy potential and the energy-momentum tensor}
\label{sec:spin+emtensor}

The knowledge of the covariant current associated to the screw torsion allows the calculation of torsional contributions
to other covariant physical quantities, most notably the spin energy potential and the energy-momentum tensor.
These are obtained by taking variations of the invariant part of the partition function~\eqref{eq:invariant_effaction} 
and isolating its local contribution~\cite{Jensen:2013kka,Manes:2018llx,Manes:2019fyw}.
In doing this, we have to keep in mind that the screw torsion is a composite field that depends not only on the torsion
tensor, 
but on the metric components as well. Thus, in computing variations with respect to the metric functions 
we should consider, in addition to the explicit dependence of the invariant action 
on these sources, also the implicit dependence in the screw torsion. 
This latter contribution can be generically written as
\begin{align}
\delta W[\mathcal{V}_{0},\mathcal{S}_{0},\boldsymbol{F}_{V},\boldsymbol{F}_{S}]_{\rm inv}&=
\int\limits_{D_{4}}\left[{\delta W_{\rm inv}\over \delta\boldsymbol{F}_{S}}d\delta\boldsymbol{S}
+{\delta W_{\rm inv}\over \delta\mathcal{S}_{0}}\delta\mathcal{S}_{0}\right] \nonumber \\[0.2cm]
&=\int\limits_{S^{3}}{\delta W_{\rm inv}\over \delta\boldsymbol{F}_{S}}\delta\boldsymbol{S}
+\int\limits_{D_{4}}\left[-d\left({\delta W_{\rm inv}\over \delta\boldsymbol{F}_{S}}\right)\delta\boldsymbol{S}
+{\delta W_{\rm inv}\over \delta\mathcal{S}_{0}}\delta\mathcal{S}_{0}\right].
\end{align}
Using the definition of the covariant current in Eq.~\eqref{eq:abelian_JScov}, we write this variation as
\begin{align}
\delta W[\mathcal{V}_{0},\mathcal{S}_{0},\boldsymbol{F}_{V},\boldsymbol{F}_{S}]_{\rm inv}
={1\over T_{0}}\int\limits_{S^{3}}\langle\boldsymbol{J}_{S}\rangle_{\rm cov}\delta\boldsymbol{S}
+\mbox{bulk terms}.
\label{eq:general_variation_covariant}
\end{align}
The previous expression can be made more explicit by writing $\langle\boldsymbol{J}_{S}\rangle_{\rm cov}$ in 
terms of its dual one-form current, defined as
\begin{align}
{1\over T_{0}}\langle\boldsymbol{J}_{S}\rangle_{\rm cov}&={1\over 2!}
\epsilon^{ijk}g_{j\ell}g_{km}\langle J_{i}\rangle_{S,\rm cov}dx^{\ell}
\wedge dx^{m}.
\end{align}
With this, Eq.~\eqref{eq:general_variation_covariant} takes the form
\begin{align}
\delta W[\mathcal{V}_{0},\mathcal{S}_{0},\boldsymbol{F}_{V},\boldsymbol{F}_{S}]_{\rm inv}&=
\int\limits_{S^{3}}d^{3}x\sqrt{g}\,g^{ij}\langle J_{i}\rangle_{S,\rm cov}\delta S_{j}.
\label{eq:general_var_inv_action}
\end{align}
This expression will be utilized next to evaluate the torsional covariant contributions to the spin energy potential and the
energy-momentum tensor.

\paragraph{The spin energy potential.}
The covariant spin energy potential is defined as~\cite{Hehl:1976kj}
\begin{align}
\delta W_{\rm inv}&=2\int\limits_{S^{3}}d^{3}x\sqrt{g}\langle \Psi_{\mu}^{\,\,\,\,\nu\sigma}\rangle_{\rm cov}
\delta T^{\mu}_{\,\,\,\nu\sigma}.
\label{eq:def_spin}
\end{align}
All dependence of the invariant effective action on the torsion tensor is included in the screw torsion field, so
we only have to care about the implicit dependence.
Using Eq.~\eqref{eq:S'svsT's}, 
we write
\begin{align}
\delta S_{m}&=-{1\over 8}e^{\sigma}\epsilon^{ijk}g_{mi}a_{\ell}\delta T^{\ell}_{\,\,\,jk}
-{1\over 4}e^{-\sigma}\epsilon^{ijk}\Big(g_{\ell j}-e^{2\sigma}a_{\ell}a_{j}\Big)\delta T^{\ell}_{\,\,\,0k}
\nonumber \\[0.2cm]
&-{1\over 8}e^{\sigma}\epsilon^{ijk}g_{mi}\delta T^{0}_{\,\,\,jk}+{1\over 4}e^{\sigma}\epsilon^{ijk}
g_{mi}a_{j}\delta T^{0}_{\,\,\,0k},
\end{align}
so the left-hand side of~\eqref{eq:def_spin} can be computed
using~\eqref{eq:general_var_inv_action}, to give
\begin{align}
\delta W_{\rm inv}
&=-{1\over 4}\int\limits_{S^{3}}d^{3}x\,\sqrt{g}e^{\sigma}\epsilon^{ijk}a_{\ell}\langle J_{i}\rangle_{S,\rm cov}
\delta T^{\ell}_{\,\,\,jk} \nonumber \\[0.2cm]
&-{1\over 4}\int\limits_{S^{3}}d^{3}x\,\sqrt{g}e^{-\sigma}
\epsilon^{ijk}\Big(g_{\ell j}-e^{2\sigma}a_{\ell}a_{j}\Big)\langle J_{i}\rangle_{S,\rm cov}
\big(\delta T^{\ell}_{\,\,\,0k}-\delta T^{\ell}_{\,\,\,k0}\big)
\\[0.2cm]
&-{1\over 4}\int\limits_{S^{3}}d^{3}x\,\sqrt{g}e^{\sigma}\epsilon^{ijk}\langle J_{i}\rangle_{S,\rm cov}\delta T^{0}_{\,\,\,jk}
+{1\over 4}\int\limits_{S^{3}}d^{3}x\,\sqrt{g}e^{\sigma}\epsilon^{ijk}
a_{j}\langle J_{i}\rangle_{S,\rm cov}\Big(\delta T^{0}_{\,\,\,0k}-\delta T^{0}_{\,\,\,k0}\Big),
\nonumber
\end{align}
plus bulk terms that we omit. This allows the identification of the covariant components of the spin energy potential in terms of the covariant
screw-torsion current as
\begin{align}
\langle\Psi_{\ell}^{\,\,\,jk}\rangle_{\rm cov}&=
-{1\over 4}e^{\sigma}\epsilon^{ijk}a_{\ell}\langle J_{i}\rangle_{S,\rm cov}, \nonumber \\[0.2cm]
\langle\Psi_{\ell}^{\,\,\,0k}\rangle_{\rm cov}&=
-{1\over 4}e^{-\sigma}
\epsilon^{ijk}\Big(g_{\ell j}-e^{2\sigma}a_{\ell}a_{j}\Big)\langle J_{i}\rangle_{S,\rm cov}, \nonumber \\[0.2cm]
\langle\Psi_{0}^{\,\,\,jk}\rangle_{\rm cov}&=
-{1\over 4}e^{\sigma}\epsilon^{ijk}\langle J_{i}\rangle_{S,\rm cov}, \\[0.2cm]
\langle\Psi_{0}^{\,\,\,0k}\rangle_{\rm cov}&={1\over 4}e^{\sigma}\epsilon^{ijk}
a_{j}\langle J_{i}\rangle_{S,\rm cov}.
\nonumber
\end{align}
Notice, however, that only one of these quantities is KK-invariant. We therefore define the physical 
KK-invariant combinations
\begin{align}
\langle\boldsymbol{\Psi}_{\ell}^{\,\,\,jk}\rangle_{\rm cov}&\equiv 
\langle\Psi_{\ell}^{\,\,\,jk}\rangle_{\rm cov}-a_{\ell}\langle\Psi_{0}^{\,\,\,jk}\rangle_{\rm cov}, \nonumber \\[0.2cm]
\langle\boldsymbol{\Psi}_{\ell}^{\,\,\,0k}\rangle_{\rm cov}&\equiv \langle\Psi_{\ell}^{\,\,\,0k}\rangle_{\rm cov}-a_{\ell}\langle\Psi_{0}^{\,\,\,0k}\rangle_{\rm cov}
+a_{j}\Big(\langle\Psi_{\ell}^{\,\,\,jk}\rangle_{\rm cov}-a_{\ell}\langle\Psi_{0}^{\,\,\,jk}\rangle_{\rm cov}\Big), 
\nonumber \\[0.2cm]
\langle\boldsymbol{\Psi}_{0}^{\,\,\,jk}\rangle_{\rm cov}&\equiv \langle\Psi_{0}^{\,\,\,jk}\rangle_{\rm cov}, \\[0.2cm]
\langle\boldsymbol{\Psi}_{0}^{\,\,\,0k}\rangle_{\rm cov}&\equiv 
\langle\Psi_{0}^{\,\,\,0k}\rangle_{\rm cov}+a_{j}\langle\Psi_{0}^{\,\,\,jk}\rangle_{\rm cov},
\nonumber
\end{align}
which are found to be
\begin{align}
\langle\boldsymbol{\Psi}_{\ell}^{\,\,\,jk}\rangle_{\rm cov}&=0, \nonumber \\[0.2cm]
\langle\boldsymbol{\Psi}_{\ell}^{\,\,\,0k}\rangle_{\rm cov}&=-{1\over 4}e^{-\sigma}\epsilon^{ijk}g_{\ell j}\langle J_{i}\rangle_{S,\rm cov}, \nonumber \\[0.2cm]
\langle\boldsymbol{\Psi}_{0}^{\,\,\,jk}\rangle_{\rm cov}&=-{1\over 4}e^{\sigma}\epsilon^{ijk}\langle J_{i}\rangle_{S,\rm cov},\\[0.2cm]
\langle\boldsymbol{\Psi}_{0}^{\,\,\,0k}\rangle_{\rm cov}&=0.
\nonumber
\end{align}
This shows that all dependence of the spin energy potentials on the torsion comes through the components of the 
covariant current associated to the screw torsion. Thus,
using Eq.~\eqref{eq:abelian_JScov}, we conclude that there are no linear torsional 
corrections to the spin energy potential. 
Notice that this quantity is nonzero even for backgrounds with vanishing torsion\footnote{Other 
contributions to the spin current in the absence of torsion have been recently computed in~\cite{Gallegos:2021bzp}.}.

\paragraph{Torsional contributions to the energy-momentum tensor.}
To evaluate the components of the (covariant) energy-momentum tensor, we take variations with respect to the 
metric functions $\sigma$, $a_{i}$ and $g_{ij}$ parametrizing the static line element~\eqref{eq:static_line_element}.
In doing this, it is very important to keep track of all dependence of the effective action on the metric.
Concerning the torsion, the definition of the $T^{\mu}$ two-form indicates that  
the components $T^{\mu}_{\,\,\,\,\nu\sigma}$ are metric-independent~\cite{Hehl:1976kj,Ortin:2015hya,Freedman:2012zz,Valle:2015hfa}. The same can be said of the 
contorsion tensor $\kappa^{\mu}_{\,\,\,[\alpha\beta]}$, while its symmetric part 
$\kappa^{\mu}_{\,\,\,(\alpha\beta)}$ depends on the metric since its computation
requires raising and lowering indices of the metric-independent torsion components 
[cf.~\eqref{eq:symmetric_part_kappa}]. 
Another important point to take into account is 
that $W_{\rm inv}$ explicitly depends on the KK vector $a_{i}$, whereas the
dependence on the other two functions comes exclusively through the screw torsion. 

We begin varying the invariant effective action with respects to~$\sigma$, which gives the components $\langle\Theta_{00}\rangle_{\rm cov}$
of the energy-momentum tensor as [cf.~\eqref{eq:general_var_inv_action}]
\begin{align}
-\int\limits_{S^{3}}
d^{3}x\sqrt{g}\,e^{-2\sigma}\langle\Theta_{00}\rangle_{\rm cov}
=\int\limits_{S^{3}}d^{3}x\sqrt{g}\,g^{ij}\langle J_{i}\rangle_{S,\rm cov}\delta_{\sigma}S_{j}.
\end{align}
Using~\eqref{eq:S'svsT's}, the variation of the screw torsion with respect to $\sigma$ appearing on the right-hand side is given by
\begin{align}
\delta_{\sigma}S_{m}&=\left(S_{m}+{1\over 2}e^{-\sigma}\epsilon^{ijk}g_{im}g_{j\ell}T^{\ell}_{\,\,\,0k}\right)\delta
\sigma,
\end{align} 
so we arrive at the result
\begin{align}
\langle\Theta_{00}\rangle_{\rm cov}&=-e^{2\sigma}
\left(S^{i}+{1\over 2}e^{-\sigma}\epsilon^{ijk}g_{j\ell}T^{\ell}_{\,\,\,0k}\right)\langle J_{i}\rangle_{S,\rm cov}.
\label{eq:theta00}
\end{align}

The mixed components $\langle\Theta_{0}^{\,\,\,i}\rangle_{\rm cov}$ are evaluated next. 
As pointed out above, in this case we have also a contribution coming from the explicit dependence of the effective action on~$a_{i}$. This 
can be computed from the invariant part of the effective action using~\cite{Manes:2018llx,Manes:2019fyw}
\begin{align}
\langle\boldsymbol{\Theta}\rangle_{\rm cov}^{\rm expl}=T_{0}\left[{\delta\over\delta(da)}-\mathcal{V}_{0}
{\delta\over\delta\boldsymbol{F}_{V}}-\mathcal{S}_{0}{\delta\over\delta\boldsymbol{F}_{S}}\right]
W[\mathcal{V}_{0},\mathcal{S}_{0},\boldsymbol{V},\boldsymbol{S},da]_{\rm inv}.
\end{align}
This is a spatial two-form, whose three-dimensional Hodge dual gives the sought components 
\begin{align}
\langle\Theta_{0}^{\,\,\,i}\rangle_{\rm cov}^{\rm expl}
&=-{1\over 4\pi^{2}}e^{-\sigma}\epsilon^{ijk}\Big(2\mathcal{S}_{0}\mathcal{V}_{0}\partial_{j}V_{k}
+\mathcal{V}_{0}^{2}\partial_{j}S_{k}+\mathcal{S}_{0}\mathcal{V}_{0}^{2}\partial_{j}a_{k}\Big) \nonumber \\[0.2cm]
&=-{1\over 4\pi^{2}}e^{\sigma}\Big(2\mu\mu_{S}\mathbb{B}^{i}+\mu^{2}B^{i}_{S}+4\mu^{2}\mu_{S}\omega^{i}\Big).
\label{eq:thetai0_exp}
\end{align}
The implicit contribution, on the other hand, is readily computed using
\begin{align}
\int\limits_{S^{3}}d^{3}x\sqrt{g}\,e^{-\sigma}\langle\Theta_{0}^{\,\,\,i}\rangle_{\rm cov}^{\rm impl}
\delta a_{i}
&=\int\limits_{S^{3}}d^{3}x\sqrt{g}\,g^{ij}\langle J_{i}\rangle_{S,\rm cov}\delta_{a}S_{j},
\end{align}
where the variation of the screw torsion with respect to the KK gauge field has the form
\begin{align}
\delta_{a}S_{m}&=-{1\over 8}\epsilon^{ijk}g_{mi}\Big(T^{\ell}_{\,\,\,jk}-2a_{j}T^{\ell}_{\,\,\,0k}\Big)
\delta a_{\ell}+{1\over 4}\epsilon^{ijk}g_{mi}\Big(T^{0}_{\,\,\,0k}+a_{\ell}T^{\ell}_{\,\,\,0k}\Big)\delta a_{j}
\nonumber \\[0.2cm]
&=-{1\over 8}g_{mi}\Big[\epsilon^{ijk}\Big(T^{\ell}_{\,\,\,jk}-2a_{j}T^{\ell}_{\,\,\,0k}\Big)
-2\epsilon^{i\ell k}\Big(T^{0}_{\,\,\,0k}+a_{j}T^{j}_{\,\,\,0k}\Big)\Big]\delta a_{\ell}.
\end{align}
This leads to
\begin{align}
\langle\Theta_{0}^{\,\,\,i}\rangle_{\rm cov}^{\rm impl}&=
-{1\over 8}e^{\sigma}\Big[\epsilon^{\ell jk}\Big(T^{i}_{\,\,\,jk}-2a_{j}T^{i}_{\,\,\,0k}\Big)
+2\epsilon^{i\ell k}\Big(T^{0}_{\,\,\,0k}+a_{j}T^{j}_{\,\,\,0k}\Big)\Big]\langle J_{\ell}\rangle_{S,\rm cov}.
\label{eq:thetai0}
\end{align}
This expression has to be added to the explicit contribution computed in Eq.~\eqref{eq:thetai0_exp}.

Finally, we evaluate the spatial components of the energy-momentum tensor
\begin{align}
{1\over 2}\int\limits_{S^{3}}d^{3}x\sqrt{g}\,\langle\Theta^{ij}\rangle_{\rm cov}
\delta g_{ij}=\int\limits_{S^{3}}d^{3}x\sqrt{g}\,g^{ij}\langle J_{i}\rangle_{\rm cov}\delta_{g}S_{j},
\end{align}
where the variation on the right-hand side of this equation takes the form
\begin{align}
\delta_{g}S_{m}&=\left(S^{k}\delta^{(i}_{k}\delta^{j)}_{m}-{1\over 2}g^{ij}S_{m}
+{1\over 4}e^{-\sigma}g_{m\ell}\epsilon^{\ell k(i}
T^{j)}_{\,\,\,\,\,0k}\right)\delta g_{ij}.
\end{align}
Thus, we arrive at
\begin{align}
\langle\Theta^{ij}\rangle_{\rm cov}&=
\left(2S^{(i}g^{j)\ell}-g^{ij}S^{\ell}+{1\over 2}e^{-\sigma}\epsilon^{\ell k(i}
T^{j)}_{\,\,\,\,\,0k}\right)\langle J_{\ell}\rangle_{S,\rm cov}.
\label{eq:thetaij}
\end{align}
The current $\langle J_{i}\rangle_{S,\rm cov}$ appearing on the right-hand side of 
Eqs.~\eqref{eq:theta00},~\eqref{eq:thetai0}, and~\eqref{eq:thetaij} is the one given in the second line
of~\eqref{eq:torsional_effects_vmodel}, which does not depend on the torsion. After
substituting the explicit expressions of the KK-invariant components of the 
axial-vector effective field given in Eq.~\eqref{eq:S'svsT's}, 
a number of new terms appears in the constitutive relations
of the energy-momentum currents which are linear in the torsion.

In section~\ref{sec:hydro_torsion} we discussed how, out of the 24 independent components of the torsion tensor~$T_{\mu\nu\alpha}$,
the microscopic Dirac field only couples to four of them contained in its antisymmetric part~$T_{[\mu\nu\alpha]}$ defining the three-form~\eqref{eq:antisymm_torsion_form}. They are 
codified in the screw torsion gauge field~$S_{\mu}$, whose components are given in terms of the torsion tensor 
in Eq.~\eqref{eq:S'svsT's}. 
The antisymmetric components of the torsion tensor are the only ones appearing in both 
the gauge currents and the spin energy potential tensor, which 
are fully written in terms of $\mathcal{S}_{0}$ and $S_{i}$. In the case of the energy-momentum tensor, however,  
an inspection of our results ~\eqref{eq:theta00},~\eqref{eq:thetai0}, and~\eqref{eq:thetaij} shows that they
contain additional components of the torsion besides the four included in the 
equilibrium partition function. In particular, the components $T_{00i}$
enter the energy-momentum tensor through
\begin{align}
T^{i}_{\,\,\,0j}=g^{ik}\Big(T_{k0j}-a_{k}T_{00j}\Big),
\end{align} 
while they cancel out from the expression of the screw torsion in~\eqref{eq:S'svsT's}.
This state of affairs is not a consequence of dimensional reduction, but a result of taking a variation with
respect to the metric components. A similar situation was found also in $2+1$~dimensions~\cite{Valle:2015hfa}.

It is known that in the presence of torsion the Ricci tensor acquires extra terms that render it nonsymmetric. This
has important consequences for the Einstein equations, which force to correct the metric energy-momentum 
tensor by a Belinfante-Rosenfeld term depending on the spin energy potential~\cite{Hehl:1976kj,Ortin:2015hya}
\begin{align}
\langle T^{\mu\nu}_{(\rm can)}\rangle_{\rm cov}=
\langle\Theta^{\mu\nu}\rangle_{\rm cov}-\stackrel{*}{\nabla}_{\alpha}\langle\Psi^{\alpha\mu\nu}\rangle_{\rm cov},
\label{eq:canon_emtensor}
\end{align}
where the modified covariant derivative is given by
\begin{align}
\stackrel{*}{\nabla}_{\mu}=\nabla_{\mu}+T^{\alpha}_{\,\,\,\,\alpha\mu}
\label{eq:modified_nabla}
\end{align}
The form of the Einstein equations is then preserved, provided its right-hand side is given by this new canonical
nonsymmetric 
energy-momentum tensor, instead of the metric one.

A relevant question here is whether the canonical energy momentum-tensor~\eqref{eq:canon_emtensor} 
plays any role at all in our analysis of
torsional fluids. In deciding this, we
should not forget that in condensed matter applications the background spacetime geometry is
nondynamical, but a way of incorporating certain microscopic features of the system into the
effective theory of the long wavelength modes. As no field equations have to be satisfied by the
background fields, 
the metric energy-momentum tensor and the spin energy potential can be regarded as two independent currents,
which provide the response to the external sources~$G_{\mu\nu}$ and~$T^{\mu}_{\,\,\,\,\alpha\beta}$.   

Finally, it should be pointed out that
although our analysis includes the linear effects of torsion, it does not incorporate 
the contribution to the partition function of the mixed gauge-gravitational anomaly in the 
background~\eqref{eq:static_line_element}. This is the reason why we can restrict our attention to the
consistent energy-momentum tensor without having to consider the gravitational BZ term.

\section{Torsional chiral effects in a two-flavor hadronic superfluid}
\label{sec:hadronic}

The results of Sec.~\ref{sec:hydro_torsion} make it easy to compute the interplay of torsion with the
dynamics of Nambu-Goldstone boson for the symmetry breaking pattern $\mbox{U(1)}_{L}\times\mbox{U(1)}_{R}
\rightarrow \mbox{U(1)}_{V}$. Our theory is characterized by three Abelian gauge fields: a vector $(\mathcal{V}_{0},
\boldsymbol{V})$ and two axial-vectors, the auxiliary $(\mathcal{A}_{0},\boldsymbol{A})$ and 
the screw torsion $(\mathcal{S}_{0},\boldsymbol{S})$.
Since the axial-vector and screw torsion fields couple in the same way to the fundamental fermions,
they always
appear in the combination $(\mathcal{A}_{0}+\mathcal{S}_{0},\boldsymbol{A}+\boldsymbol{S})$ in the effective
action. Setting $\boldsymbol{A}=0$ at the end of the calculation, 
the linear couplings of the single Nambu-Goldstone boson $\alpha$ in the
 local WZW action
takes a particularly simple form~\cite{Manes:2018llx}
\begin{align}
W[\alpha,\ldots]_{\rm WZW}&\equiv \Big.\Big(
W[\ldots,\boldsymbol{A}+\boldsymbol{S},\ldots]_{\rm anom}-W[\ldots,\boldsymbol{A}+
\boldsymbol{S}+d\alpha,\ldots]_{\rm anom} \Big)\Big|_{\boldsymbol{A}=0}
\nonumber\\[0.2cm]
&=\ldots
-{1\over 6\pi^{2}T_{0}}\int\limits_{S^{3}}\mathcal{A}_{0}\Big(\boldsymbol{F}_{S}+\mathcal{S}_{0}da\Big)d\alpha,
\end{align}
where the ellipsis in the second line 
indicates the torsion-independent coupling of the Nambu-Goldstone boson to the background fields
$\mathcal{V}_{0}$, $\boldsymbol{V}$, and $\mathcal{A}_{0}$. We see that the Nambu-Goldstone boson couples linearly to 
$\boldsymbol{B}_{S}$,
the magnetic component of the screw torsion field strength defined in Eq.~\eqref{eq:emdecomFS}.
Expressed in components, the relevant term in the WZW action reads
\begin{align}
W[\alpha,\ldots]_{\rm WZW}&=\ldots
-\int\limits_{S^{3}}d^{3}x\,\sqrt{g}\,{\mu_{5}\over 6\pi^{2}T}B_{S}^{i}\partial_{i}\alpha,
\end{align}
where $B_{S}^{i}$ was defined in Eq.~\eqref{eq:BsDef}
and we also introduced the local temperature $T=e^{-\sigma}T_{0}$ and the chiral chemical potential $\mu_{5}=e^{-\sigma}
\mathcal{A}_{0}$.

After this brief discussion of the Abelian case, 
we turn our attention to the analysis of torsional chiral effects in the 
two-flavor hadronic superfluid studied in Ref.~\cite{Manes:2019fyw}, where we refer the reader for details. 
This fluid couples to a vector and an axial-vector external gauge fields,
respectively denoted by $\mathcal{V}$ and $\mathcal{A}$,
transforming in the flavor group $\mbox{U(2)$_{L}\times$U(2)$_{R}$}$. The values of these background fields are restricted to the Cartan subalgebra 
generated by
\begin{align}
t_{0}={1\over 2}\mathbb{1}, \hspace*{1cm} t_{3}={1\over 2}\sigma_{3}.
\end{align}
We assume the system undergoes spontaneous symmetry breaking to its vector subgroups, $\mbox{U(2)$_{L}\times$U(2)$_{R}$}\rightarrow 
\mbox{U(1)$_{V}$}\times\mbox{SU(2)$_{V}$}$, generating in the process a triplet of Nambu-Goldstone bosons $\pi^{0},\pi^{\pm}$ encoded in the
matrix
\begin{align}
U=\exp\left[{i\sqrt{2}\over f_\pi} 
\left(
\begin{array}{cc}
\tfrac{1}{\sqrt{2}} \pi^0 & \pi^+ \\
\pi^- & -\tfrac{1}{\sqrt{2}} \pi^0
\end{array} \right)\right].
\end{align}

The first issue to address is how to incorporate the torsional vector fields into our analysis. 
The left Dirac operator, including the vector, axial, and screw-torsion fields, takes the form 
\begin{align}
\overrightarrow{D\!\!\!\!/}\,\psi&=\Big(\overrightarrow{\fslashnablabar}-i{\mathcal{V}\!\!\!\!/\,}-i{\mathcal{A}\!\!\!\!/\,}\gamma_{5}-i\fslash{\mathcal{S}}\mathbb{1}\gamma_{5}\Big)\psi \nonumber \\[0.2cm]
&=\Big\{\overrightarrow{\fslashnablabar}-i\Big({\mathcal{V}\!\!\!\!/\,}_{0}t_{0}+{\mathcal{V}\!\!\!\!/\,}_{3}t_{3}\Big)
-i\Big[\big({\mathcal{A}\!\!\!/\,}_{0}+2\,\fslash{\mathcal{S}}\big)t_{0}+{\mathcal{A}\!\!\!/\,}_{3}t_{3}\Big]\gamma_{5}
\Big\}\psi,
\end{align}
and similarly for the right operator. 
The structure of these terms shows that, to take into account the effect of torsion in the analysis 
of Refs.~\cite{Manes:2019fyw}, it is enough to implement the 
replacement
\begin{align}
\mathcal{A}_{0\mu}&\longrightarrow \mathcal{A}_{0\mu}+2\mathcal{S}_{\mu},
\end{align}
while leaving all remaining fields unchanged.  

With all this in mind, we can compute the linear torsional
corrections to the covariant gauge currents at leading order in the derivative 
expansion. To avoid cumbersome expressions, here we only give the 
terms in the currents depending linearly on the torsion, that we denote by
$\langle \Delta\mathcal{J}^{\mu}_{aV}\rangle_{\rm cov}$ and
$\langle \Delta\mathcal{J}^{\mu}_{aA}\rangle_{\rm cov}$. These should be added to the expressions found in Ref.~\cite{Manes:2019fyw} for
the corresponding currents. The results are
\begin{align}
\langle \Delta\mathcal{J}^{\mu}_{0V}\rangle_{\rm cov}&=-{N_{c}\over 8\pi^{2}}\epsilon^{\mu\nu\alpha\beta}
\mathcal{S}_{\nu}\mathcal{V}_{0\alpha\beta}, \nonumber \\[0.2cm]
\langle \Delta\mathcal{J}_{3V}^{\mu}\rangle_{\rm cov}&=-{N_{c}\over 24\pi^{2}}\epsilon^{\mu\nu\alpha\beta}\Big[
\big(\mathds{H}+3\big)\mathcal{S}_{\nu}\mathcal{V}_{3\alpha\beta}-\partial_{\alpha}\big(\mathcal{S}_{\nu}\mathds{T}_{\beta}\big)\Big], \nonumber\\[0.2cm]
\langle \Delta \mathcal{J}^{\mu}_{0A}\rangle_{\rm cov}&=-{N_{c}\over 24\pi^{2}}\epsilon^{\mu\nu\alpha\beta}
\Big(\mathcal{S}_{\nu}\mathcal{A}_{0\alpha\beta}+2\mathcal{A}_{0\nu}\partial_{\alpha}\mathcal{S}_{\beta}\Big),
 \\[0.2cm]
\langle \Delta\mathcal{J}^{\mu}_{3A}\rangle_{\rm cov}&={N_{c}\over 24\pi^{2}}\epsilon^{\mu\nu\alpha\beta}\partial_{\nu}\big(\mathcal{S}_{\alpha}\mathds{I}_{\beta}\big),
\nonumber
\end{align}
where we have used the tensor structures introduced in~\cite{Manes:2019fyw}
\begin{align}
\mathds{H}&\equiv {\rm Tr\,}\Big[\Big(U^{-1}QU-Q\Big)Q\Big], \nonumber \\[0.2cm]
\mathds{I}_{\mu}&\equiv {\rm Tr\,}\Big[\Big(R_{\mu}+L_{\mu}\Big)Q\Big], \label{eq:HIT}\\[0.2cm]
\mathds{T}_{\mu}&\equiv {\rm Tr\,}\Big[Q\Big(R_{\mu}-L_{\mu}\Big)\Big]+2\mathcal{V}_{3\mu}{\rm Tr\,}\Big[\Big(U^{-1}QU-Q\Big)Q\Big],
\nonumber
\end{align}
with $R_{\mu}=iU^{-1}\partial_{\mu}U$ and $L_{\mu}=i\partial_{\mu}UU^{-1}$, while the charge matrix
is given by
\begin{align}
Q={1\over 3}t_{0}+t_{3}.
\label{eq:charge_matrix_def}
\end{align}
The first thing to be noticed here is that
the torsional couplings of the pions are restricted to the three-flavor components of the vector and axial-vector covariant currents. 

In order to compute the longitudinal and transverse components of these currents,
and write the constitutive relations 
of the hadronic superfluid, 
we need to introduce a number of scalar and vector structures,
in addition to those used in Ref.~\cite{Manes:2019fyw}\footnote{For the reader's convenience, these are summarized in Appendix \ref{app:identities}.}. 
They represent the linear coupling of the Nambu-Goldstone bosons to torsion. To the five scalar ones, we add
\begin{align}
\mathds{S}_{6}&= \epsilon^{\mu\nu\alpha\beta}u_{\mu}\partial_{\nu}\big(\mathcal{S}_{\alpha}\mathds{I}_{\beta}\big), \nonumber \\[0.2cm]
\mathds{S}_{7}&= \epsilon^{\mu\nu\alpha\beta}u_{\mu}\partial_{\nu}\big(\mathcal{S}_{\alpha}\mathds{T}_{\beta}\big), 
\end{align}
while for the vector structures we extend the notation $\mathds{P}^{\mu}_{1,a}$ and $\mathds{P}^{\mu}_{3,a}$
to include $a=S$, and add a new term $\mathds{P}^{\mu}_{5}$
\begin{align}
\mathds{P}_{1,S}^{\mu}&=\epsilon^{\mu\nu\alpha\beta}u_{\nu}\mathds{I}_{\alpha}\partial_{\beta}\left({\mu_{S}\over T}\right), \nonumber \\[0.2cm]
\mathds{P}_{3,S}^{\mu}&=\epsilon^{\mu\nu\alpha\beta}u_{\nu}\mathds{T}_{\alpha}\partial_{\beta}\left({\mu_{S}\over T}\right), \\[0.2cm]
\mathds{P}_{5}^{\mu}&= \epsilon^{\mu\nu\alpha\beta}u_{\nu}\mathcal{S}_{\alpha}\partial_{\beta}\mathds{H}.
\nonumber
\end{align}
Here we have used the screw torsion chemical potential~\eqref{eq:chemical_potentials}, as well as the local temperature $T=e^{-\sigma}T_{0}$.

The longitudinal and transverse components of the covariant vector and axial-vector 
currents are now written in terms of these quantities. The
new torsional nondissipative chiral transport coefficients can be read from the resulting expressions. 
We start with the 0-flavor vector current
\begin{align}
u_{\mu}\langle \Delta\mathcal{J}^{\mu}_{0V}\rangle_{\rm cov}&={N_{c}\over 4\pi^{2}}\mathcal{S}_{\mu}\mathcal{B}_{0}^{\mu}, \nonumber \\[0.2cm]
P^{\mu}_{\,\,\,\sigma}\langle\Delta\mathcal{J}^{\sigma}_{0V}\rangle_{\rm cov}&=
{N_{c}\over 4\pi^{2}}\left[\mu_{S}\mathcal{B}^{\mu}_{0}+
T\epsilon^{\mu\nu\alpha\beta}u_{\nu}\mathcal{S}_{\alpha}\partial_{\beta}\left({\mu_{0}\over T}\right)\right],
\end{align}
where we have introduced the magnetic field
\begin{align}
\mathcal{B}_{a}^{\mu}={1\over 2}\epsilon^{\mu\nu\alpha\beta}u_{\nu}\mathcal{V}_{a\alpha\beta} \hspace*{2cm} a=0,3.
\label{eq:magnetic_field_def}
\end{align}
The longitudinal and transverse components of the 3-flavor vector current, on the other hand, read
\begin{align}
u_{\mu}\langle \Delta\mathcal{J}_{3V}^{\mu}\rangle_{\rm cov}&=
{N_{c}\over 24\pi^{2}}\Big[
2\big(\mathds{H}+3\big)\mathcal{S}_{\mu}\mathcal{B}_{3}^{\mu}
-\mathds{S}_{7}\Big],
\nonumber \\[0.2cm]
P^{\mu}_{\,\,\,\sigma}\langle \Delta\mathcal{J}_{3V}^{\sigma}\rangle_{\rm cov}&=
{N_{c}\over 24\pi^{2}}\bigg[2\big(\mathds{H}+3\big)\mu_{S}\mathcal{B}^{\mu}_{3}+\big(\mathds{H}+6\big)
T\epsilon^{\mu\nu\alpha\beta}u_{\nu}\mathcal{S}_{\alpha}\partial_{\beta}\left({\mu_{3}\over T}\right) \nonumber \\[0.2cm]
&-\mu_{S}\mathds{P}_{4}^{\mu}
+T\mathds{P}_{3,S}^{\mu}
+\mu_{3}\mathds{H}\epsilon^{\mu\nu\alpha\beta}u_{\nu}\partial_{\alpha}\mathcal{S}_{\beta}
-\mu_{3}\mathds{P}_{5}^{\mu}\bigg].
\end{align}
In the case of the axial-vector currents, we have
\begin{align}
u_{\mu}\langle \Delta \mathcal{J}^{\mu}_{0A}\rangle_{\rm cov}&=
{N_{c}\mu_{5}\over 6\pi^{2}}\mathcal{S}_{\mu}\omega^{\mu}, \nonumber \\[0.2cm]
P^{\mu}_{\,\,\,\sigma}\langle \Delta \mathcal{J}^{\sigma}_{0A}\rangle_{\rm cov}&=
{N_{c}\mu_{5}\over 12\pi^{2}}\bigg(-2\mu_{S}\omega^{\mu}
+\epsilon^{\mu\nu\alpha\beta}u_{\nu}\partial_{\alpha}\mathcal{S}_{\beta}\bigg),
\end{align}
for the 0-flavor components, whereas in the case of the 3-flavor the result is
\begin{align}
u_{\mu}\langle \Delta\mathcal{J}^{\mu}_{3A}\rangle_{\rm cov}&={N_{c}\over 24\pi^{2}}\mathds{S}_{6}, \nonumber \\[0.2cm]
P^{\mu}_{\,\,\,\sigma}\langle \Delta\mathcal{J}^{\sigma}_{3A}\rangle_{\rm cov}
&=-{N_{c}\over 24\pi^{2}}\Big(\mu_{S}\mathds{P}_{2}^{\mu}
-T\mathds{P}_{1,S}^{\mu}\Big).
\end{align}
Let us stress once more that all the expressions given here only represent the linear torsional contributions to the longitudinal and
transverse components of the covariant currents, 
that should be added to the 
respective nontorsional terms found in~\cite{Manes:2019fyw}.
As already pointed out, 
torsion couples to Nambu-Goldstone bosons only through the 3-flavor currents.
The torsional contributions to the 0-flavor covariant vector and axial-vector currents are 
just given by the corresponding terms of the BZ currents linear in the torsion. This implies that 
there are no linear torsional contributions to the 0-flavor component of the consistent currents and, as a consequence, 
no linear torsional terms in the WZW action depending on~$V_{0i}$ or~$A_{0i}$. 

To find the expression of the covariant currents in terms of the physical electromagnetic fields, 
we expand the KK-invariant components of the vector field in terms of the charge matrix $Q$ defined 
in~\eqref{eq:charge_matrix_def} and the generator
$t_{3}$ according to (see~\cite{Manes:2019fyw})
\begin{align}
V_{0\mu}t_{0}+V_{3\mu}t_{3}=3V_{0\mu}Q+\big(V_{3\mu}-3V_{0\mu}\big)t_{3}.
\end{align}
As usual, the unbroken U(1)$_{\rm V}$ factor is identified as the one coupling to the $Q$ matrix, whereas the
field coupling to $t_{3}$ is set to zero, which implies
$V_{\mu}\equiv 3V_{0\mu}=V_{3\mu}$.
We write now the torsional terms in the covariant electromagnetic 
current
\begin{align}
\langle\mathcal{J}^{\mu}_{\rm em}\rangle_{\rm cov}={e\over 3}\langle\mathcal{J}_{0V}^{\mu}\rangle_{\rm cov}
+e\langle\mathcal{J}^{\mu}_{3V}\rangle_{\rm cov},
\end{align}
in terms of the pion fields as 
\begin{align}
\langle \Delta\mathcal{J}^{i}_{\rm em}\rangle_{\rm cov}
&=-{e^{2}N_{c}\over 12\pi^{2}f_{\pi}^{2}}\epsilon^{ijk}S_{j}\mathbb{E}_{k}\pi^{+}\pi^{-}
-{\mu e^{2}N_{c}\over 12\pi^{2}f_{\pi}^{2}}B_{S}^{i}\pi^{+}\pi^{-}
\nonumber\\[0.2cm]
&-{ieN_{c}\over 12\pi^{2}f_{\pi}^{2}}T\epsilon^{ijk}\partial_{k}\left[{\mu_{S}\over T}
\Big(\pi^{+}\partial_{j}\pi^{-}-\pi^{-}\partial_{j}\pi^{+}\Big)\right]
-{e^{2}N_{c}\over 6\pi^{2}f_{\pi}^{2}}T\epsilon^{ijk}V_{j}\partial_{k}\Big({\mu_{S}\over T}\pi^{+}\pi^{-}\Big)
\nonumber \\[0.2cm]
&+{\mu e^{2}N_{c}\over 12\pi^{2}f_{\pi}^{2}}\epsilon^{ijk}S_{j}\partial_{k}\big(\pi^{+}\pi^{-}\big)
+{5e^{2}N_{c}\over 18\pi^{2}}\Big(\mu_{S}\mathbb{B}^{i}+\epsilon^{ijk}S_{j}\mathbb{E}_{k}\Big)
+\mathcal{O}(\pi^{3}),
\label{eq:electric_current} 
\end{align}
where the electric and magnetic fields
are defined in Eq.~\eqref{eq:electric_magnetic_fields_def} and $B_{S}^{i}$ is given in~\eqref{eq:BsDef}, all
fields here being KK-invariant. The last, pion-independent term is 
the BZ electromagnetic current of the unbroken theory, and replicates the structure found in 
Eq.~\eqref{eq:BZemcurrent} for the Abelian case. We see that 
there is no torsion-mediated electromagnetic coupling to the neutral pion. There exits nonetheless a torsional pion-dependent contribution to the
chiral electric effect given by the first term in Eq.~\eqref{eq:electric_current}, this time induced by the $\mathsf{T}$-odd spatial screw-torsion field. 

As for the transverse axial-vector currents, we see that the only coupling of torsion to pions arises from the 3-flavor component
\begin{align}
\langle \Delta\mathcal{J}^{i}_{3A}\rangle_{\rm cov}
&=-{N_{c}\over 12\pi^{2}f_{\pi}}T\epsilon^{ijk}\partial_{j}\pi^{0}\partial_{k}\left({\mu_{S}\over T}\right)+\mathcal{O}(\pi^{3}).
\label{eq:J3A}
\end{align} 
There is therefore no torsional corrections to the pion-mediated chiral electric, magnetic, and vortical separation effects
found in~\cite{Manes:2019fyw}. Interestingly, 
the coupling showed in Eq.~\eqref{eq:J3A} is the only torsion-induced term involving the neutral pion in the 
constitutive relations at this order.

\section{Closing remarks}
\label{sec:conclusions}

In this paper we have analyzed the linear effect of background torsion in the partition function of a charged fluid 
minimally coupled to gravity. The terms studied are those induced by the 't Hooft anomaly associated with the
screw torsion, dual to the antisymmetric part of the torsion tensor. In the Abelian case,
our results show the existence of magnetic and vortical chiral torsional effects, whereas the 
axial-vector current does not have any linear torsional corrections.

In this same model, the covariant spin energy potential and energy-momentum tensor have been computed in terms of the
axial-vector covariant current, with coefficients that only depend on the metric functions. Since the axial-vector current
has been shown not to depend on the torsion at linear order, we conclude that the spin energy potential does not exhibit
linear torsional contributions. The situation is quite different in the case of the covariant energy-momentum tensor.
Although its components are also written in terms of the covariant axial-vector currents, the coefficients now
do depend linearly on the torsion tensor. Thus, we find linear torsional corrections to the energy-momentum tensor, 
which actually include components of the torsion tensor that do not appear in the effective action.
This latter situation is analogous to the one already found in $2+1$ dimension~\cite{Valle:2015hfa}.
It is important to stress that the torsional contributions to the energy-momentum tensor and the spin
energy potential are associated with the implicit dependence of the effective axial-vector gauge field on 
both the metric and the torsion components. This is what makes the torsional theory genuinely different from
the theory of a Dirac fermion axially coupled to an external 
gauge field, as it is reflected in the constitutive 
relations.

We have also studied linear 
torsional chiral effects in a two-flavor hadronic superfluids studied in Ref.~\cite{Manes:2019fyw}.
We found that no new couplings of the Nambu-Goldstone bosons to torsion emerge from 
the 0-flavor components of the covariant vector and axial-vector currents. 
The analysis of the electromagnetic transverse current,
on the other hand, shows the existence of torsional chiral electric effect mediated by the two charged
Nambu-Goldstone bosons $\pi^{\pm}$, 
whereas no torsional vortical effect appears. Interstingly, there are no torsional corrections to the pion-mediated
electric, magnetic, and vortical chiral separation effects that were found in~\cite{Manes:2019fyw}. 

In this paper we have exploited the analogy between background torsion and axial-vector couplings to compute the linear 
effects of torsion, which come from triangle diagrams with an axial-vector current coupled to the background screw torsion field. A different sector is the one associated with the Nieh-Yan anomaly~\cite{Nieh:1981ww,Nieh:1981xk}, 
which explicitly depends on the UV cutoff scale of the theory. This Nieh-Yan term has been shown to be relevant 
in condensed matter, where this cutoff arises naturally. It would be interesting to further explore the physical implications of this anomaly along the lines followed in the present paper for the triangle contributions. This issue 
will be addressed elsewhere.

\acknowledgments
We thank Karl Landsteiner for discussions.
This work has been supported by 
Spanish Science Ministry grants PGC2018-094626-B-C21 (MCIU/AEI/FEDER, EU) and PGC2018-094626-B-C22
(MCIU/AEI/FEDER, EU), as well as by Basque Government
grant IT979-16. 

\appendix

\section{Basics of spacetime torsion}
\label{sec:torsion}

In this appendix, we give a brief overview of the main mathematical features of spacetimes with torsion. 
The focus will lie
on the basic differential geometric aspects, with further details being available in a number of reviews (see, for example, 
\cite{Hehl:1976kj,Shapiro:2001rz,Ortin:2015hya,Trautman2006,Hehl:2007bn,Freedman:2012zz}).
Let us consider a four-dimensional curved manifold and an orthonormal tetrad basis~$\{e^{A}=e^{A}_{\,\,\,\mu}dx^{\mu}\}$
\begin{align}
\eta_{AB}e^{A}_{\,\,\,\mu}e^{B}_{\,\,\,\nu}=g_{\mu\nu},
\end{align}
with $\eta_{AB}$ the flat Lorentz metric and $g_{\mu\nu}$ the spacetime metric\footnote{In this work, Lorentz indices are denoted by 
capital Latin letters, while spacetime indices are indicated by Greek letters. Lowercase Latin indices are reserved for
spatial components. To make notation lighter, we omit the wedge~$(\wedge)$ to denote exterior products.}. 
The spin connection~$\omega^{A}_{\,\,\,B}$ defines the notion of parallel transport, allowing the construction of the 
covariant derivative operator, that in the particular case
of $p$-form tensor of the type $\tau^{A}_{\,\,\,B}$ takes the form
\begin{align}
\nabla \tau^{A}_{\,\,\,B}=d\tau^{A}_{\,\,\,B}+\omega^{A}_{\,\,\,C}\tau^{C}_{\,\,\,B}+(-1)^{p+1}\tau^{A}_{\,\,\,C}\omega^{C}_{\,\,\,B}.
\label{eq:cov_diff_def}
\end{align}
In what follows, we assume the connection $\omega^{A}_{\,\,\,BC}$ to be metric compatible, $\nabla\eta_{AB}=0$.
Torsion is defined by the first Cartan structure equation
\begin{align}
T^{A}=de^{A}+\omega^{A}_{\,\,\,B}e^{B},
\label{eq:torsion_2form_def}
\end{align}
while the second one gives the curvature
\begin{align}
R^{A}_{\,\,\,B}&=d\omega^{A}_{\,\,\,B}+\omega^{A}_{\,\,\,C}\omega^{C}_{\,\,\,B}.
\end{align}
Both torsion and curvature are geometrical quantities related to the behavior of vectors under (infinitesimal) parallel transport. 

The torsion two-form can be expanded in the tetrad basis as
\begin{align}
T^{A}={1\over 2}T^{A}_{\,\,\,BC}e^{B}e^{C},
\end{align} 
where the components on the right-hand side are antisymmetric in the lower two indices, $T^{A}_{\,\,\,(BC)}=0$. 
To parametrize torsion, it is convenient to introduce
an auxiliary torsionless connection $\overline{\omega}^{A}_{\,\,\,B}$
associated with the same tetrad basis $e^{A}$ and satisfying
\begin{align}
de^{A}+\overline{\omega}^{A}_{\,\,\,B}e^{B}=0.
\label{eq:torsionless_conn_def}
\end{align}
This auxiliary connection is also metric compatible, $\overline{\nabla}\eta_{AB}=0$, 
where here and elsewhere in the paper we indicate all quantities associated with this Levi-Civita connection by an overline. 
The contorsion one-form is defined by
\begin{align}
\kappa^{A}_{\,\,\,B}\equiv \omega^{A}_{\,\,\,B}-\overline{\omega}^{A}_{\,\,\,B},
\label{eq:contorsio_tensor_def}
\end{align}
which, being the difference of two connections, transforms as a tensor under local Lorentz transformations. Combining 
Eqs.~\eqref{eq:torsion_2form_def} and \eqref{eq:torsionless_conn_def}, we write the torsion two-form $T^{A}$ in terms of the 
contortion one-form as 
\begin{align}
T^{A}=\kappa^{A}_{\,\,\,\,\,B}e^{B}=-\kappa^{A}_{\,\,\,BC}e^{B}e^{C},
\end{align}
where in the second equality we have expanded $\kappa^{A}_{\,\,\,B}=\kappa^{A}_{\,\,\,BC}e^{C}$.
This identity shows that the antisymmetric part of the contorsion 
in the two lower indices is determined by the components of the torsion tensor
\begin{align}
\kappa^{A}_{\,\,\,[BC]}=-{1\over 2}T^{A}_{\,\,\,\,\,BC}.
\label{eq:antisymmetric_part_kappa}
\end{align}
Using metric compatibility, the symmetric piece can be computed in terms of the torsion components as
\begin{align}
\kappa^{A}_{\,\,\,(BC)}={1\over 2}T^{\,\,\,\,A}_{B\,\,\,\,C}+{1\over 2}T^{\,\,\,\,A}_{C\,\,\,\,B}.
\label{eq:symmetric_part_kappa}
\end{align}
Similar expressions are obtained using a coordinate basis, with the torsion being identified
with the antisymmetric part of the connection according to
\begin{align}
T^{\mu}_{\,\,\,\,\nu\sigma}=-2\Gamma^{\mu}_{\,\,\,[\nu\sigma]}=-2\kappa^{\mu}_{\,\,\,[\nu\sigma]}.
\label{eq:kappavsT}
\end{align}

\section{A summary of expressions 
from Ref.~\cite{Manes:2019fyw}
}
\label{app:identities}

For the sake of completeness, we list in this Appendix the scalar and 
tensor structures introduced in Ref.~\cite{Manes:2019fyw} to write the constitutive relations for the two-flavor
chiral hadronic superfluid studied in Section~\ref{sec:hadronic}. 
In the case of the longitudinal components of the currents, these are expressed in terms of the following five scalar 
structures
\begin{align}
\mathds{S}_{1,a}&\equiv \epsilon^{\mu\nu\alpha\beta}\mathds{I}_{\mu}u_{\nu}\partial_{\alpha}\mcV_{a\beta}
=\mathds{I}_{\mu}\mcB^{\mu}_{a} \hspace*{2cm}
(a=0,3), \nonumber \\[0.2cm]
\mathds{S}_{2}&\equiv {1\over 2}\epsilon^{\mu\nu\alpha\beta}\mathds{I}_{\mu}u_{\nu}\partial_{\alpha}u_{\beta}
=\mathds{I}_{\mu}\omega^{\mu}, \nonumber \\[0.2cm]
\mathds{S}_{3}&\equiv \epsilon^{\mu\nu\alpha\beta}u_{\mu}\left[\mcV_{3\nu}\partial_{\alpha}\mathds{I}_{\beta}
-{i\over 3}{\rm Tr\,}\Big(L_{\nu}L_{\alpha}L_{\beta}\Big)\right], 
\label{eq:pseudoscalarsSs}\\[0.2cm]
\mathds{S}_{4,a}&\equiv \epsilon^{\mu\nu\alpha\beta}\mathds{T}_{\mu}u_{\nu}\partial_{\alpha}\mcV_{a\beta}=\mathds{T}_{\mu}\mcB_{a}^{\mu}
\hspace*{2cm} (a=0,3),  \nonumber \\[0.2cm]
\mathds{S}_{5}&\equiv {1\over 2}\epsilon^{\mu\nu\alpha\beta}\mathds{T}_{\mu}u_{\nu}\partial_{\alpha}u_{\beta}=\mathds{T}_{\mu}\omega^{\mu}, \nonumber
\end{align}
where the magnetic field is defined in Eq.~\eqref{eq:magnetic_field_def}, and $\mathds{I}_{\mu}$ and $\mathds{T}_{\mu}$ are given in Eq.~\eqref{eq:HIT}, whose expansions in terms of the pion fields
are given by
\begin{align}
\mathds{H}&=-{2\over f_{\pi}^{2}}\pi^{+}\pi^{-}+\mathcal{O}(\pi^{4}), \nonumber \\[0.2cm]
\mathds{I}_{\mu}&=-{2\over f_{\pi}}\partial_{\mu}\pi^{0}+\mathcal{O}(\pi^{3}), 
\label{eq:HIT_pions}\\[0.2cm]
\mathds{T}_{\mu}&={2i\over f_{\pi}^{2}}\Big(\pi^{+}\partial_{\mu}\pi^{-}-\pi^{-}\partial_{\mu}\pi^{+}\Big)-
{4\over f_{\pi}^{2}}\pi^{+}\pi^{-}\mathcal{V}_{3\mu}+\mathcal{O}(\pi^{3}).
\nonumber
\end{align}
Finally, the transverse components of the covariant currents found in~\cite{Manes:2019fyw} are expressed in terms of the four tensor structures
\begin{align}
\mathds{P}^{\mu}_{1,a}&\equiv \epsilon^{\mu\nu\alpha\beta}u_{\nu}\mathds{I}_{\alpha}\partial_{\beta}\left({\mu_{a}\over T}\right) \hspace*{2cm}
(a=0,3),
\nonumber \\[0.2cm]
\mathds{P}^{\mu}_{2}&\equiv \epsilon^{\mu\nu\alpha\beta}u_{\nu}\partial_{\alpha}\mathds{I}_{\beta}, \nonumber \\[0.2cm]
\mathds{P}^{\mu}_{3,a}&\equiv \epsilon^{\mu\nu\alpha\beta}u_{\nu}\mathds{T}_{\alpha}\partial_{\beta}\left({\mu_{a}\over T}\right) \hspace*{2cm}
(a=0,3), \label{eq:pseudovectorPs}\\[0.2cm]
\mathds{P}^{\mu}_{4}&\equiv \epsilon^{\mu\nu\alpha\beta}u_{\nu}\partial_{\alpha}\mathds{T}_{\beta}.
\nonumber 
\end{align} 
These expressions have been written in terms of the chemical potentials
\begin{align}
\mu_{a}&=e^{-\sigma}\mathcal{V}_{a0}.
\end{align}

\bibliographystyle{JHEP}
\bibliography{biblio_file}

\end{document}